\documentclass[10pt,conference]{IEEEtran}
\IEEEoverridecommandlockouts

\AtBeginDocument{%
  \providecommand\BibTeX{{%
    \normalfont B\kern-0.5em{\scshape i\kern-0.25em b}\kern-0.8em\TeX}}}

\usepackage[hidelinks]{hyperref}
\usepackage{xcolor}
\usepackage{array}
\usepackage{cleveref}
\crefname{section}{\S}{\S\S}
\crefname{subsection}{\S}{\S\S}
\crefname{listing}{listing}{listings}

\usepackage[nolist]{acronym}

\usepackage{subcaption}
\usepackage{courier}
\usepackage{pxfonts}

\usepackage{listings}

\usepackage{graphicx}
\usepackage[square,sort,comma,numbers]{natbib}
\DeclareCaptionFormat{listing} {
  \parbox{\textwidth}{\hspace{-0.2cm}#1#2#3}
}

\newtheorem{example}{Example}

\newtheorem{resq}{RQ}

\usepackage{multicol}
\usepackage{enumitem}
\setitemize{noitemsep,topsep=0pt,parsep=0pt,partopsep=0pt}

\def\BibTeX{{\rm B\kern-.05em{\sc i\kern-.025em b}\kern-.08em
    T\kern-.1667em\lower.7ex\hbox{E}\kern-.125emX}}
\begin{document}

\lstset{ 
  language=Matlab, 
  tabsize=2, 
  showspaces=false, 
  showstringspaces=false, 
  float=[htb], 
  basicstyle=\footnotesize, 
  frame=tbrl, 
  frameround=tttt, 
  numbers=left, 
  numberstyle=\tiny, 
  numberblanklines=false, 
  linewidth=.4\textwidth
}

\title{GeNet: A Multimodal LLM-Based Co-Pilot for
Network Topology and Configuration}

\author{
    \IEEEauthorblockN{
        Beni Ifland\IEEEauthorrefmark{1},
        Rubin Krief\IEEEauthorrefmark{1},
        Aviram Zilberman\IEEEauthorrefmark{1},
        Elad Duani\IEEEauthorrefmark{1},
        Miro Ohana\IEEEauthorrefmark{1},
        Andres Murillo\IEEEauthorrefmark{2},\\
        Ofir Manor\IEEEauthorrefmark{2},
        Ortal Lavi\IEEEauthorrefmark{2},
        Kenji Hikichi\IEEEauthorrefmark{2},
        Asaf Shabtai\IEEEauthorrefmark{1},
        Yuval Elovici\IEEEauthorrefmark{1},
        Rami Puzis\IEEEauthorrefmark{1}
    }

    \IEEEauthorblockA{\IEEEauthorrefmark{1}Ben Gurion University of the Negev, Software and Information Systems Engineering and Cyber@BGU \\
    }

    \IEEEauthorblockA{\IEEEauthorrefmark{2}Fujitsu}
}
\maketitle
\begin{abstract}
 Managing communication networks in enterprise environments is complex, time-consuming, and error-prone. 
 Research on automating network engineering has mainly focused on configuration synthesis, while changes to the physical network topology are often overlooked. 
 In this paper, we introduce \textit{GeNet}, which combines visual and textual information to interpret and modify network topologies and device configurations based on user intents. 
 We evaluated \textit{GeNet} using scenarios adapted from Cisco certification exercises relevant to enterprise networks and a series of different topology image variants. 
 Our results show that \textit{GeNet} can interpret images of network topologies, including low-quality handwritten ones, which can reduce the workload of network engineers and accelerate the network design process.
 Moreover, \textit{GeNet} demonstrates the ability to handle intents successfully, even with incomplete or varying quality input specifications.
 
\end{abstract}


\vspace{-1em}
\section{Introduction}\label{sec:intro}

Communication network engineering in enterprise environments is a complex, time-consuming, and error-prone process that relies heavily on large IT, NetOps, and network engineering teams. 
As modern enterprise network architectures have become more diverse and complex, these tasks have emerged as even more challenging~\cite{houidi2022neural}. 
Researchers from academia and industry have attempted to simplify and automate network engineering workflows. One notable approach is \acf{IBN}~\cite{leivadeas2022survey} which aims to serve as a bridge between high-level network requirements, commonly referred to as intents, and low-level implementations. Originally, \ac{IBN} frameworks relied on formal languages to express intents and provided tools for optimizing network configurations to satisfy them~\cite{beckett2016don}. 
Later approaches incorporated \ac{NLP}-based methods to reduce the formality, further assisting network engineers~\cite{wang2024netconfeval,jacobs2021hey}.

Inspired by advances in the use of co-pilots for programming, such as GitHub Copilot\footnote{https://github.com/features/copilot} 
researchers are examining how co-pilots and recommendation systems can be tailored to assist network engineers in network device configuration~\cite{guo2023configreco,zhao2023confpilot}. 
However, despite recent advances, much of the research on network engineering automation has mainly focused on configuration synthesis, often overlooking the essential task of network topology design. 
Engineers testify that whiteboards, though imperfect, are vital for network design\footnote{https://rule11.tech/the-white-board-and-the-simulation/}.
Moreover, Wang et al.~\cite{wang2024netconfeval} suggested that including information such as network topology images, could provide valuable context for generating network configurations.
We streamline \ac{IBN} processes by enabling topology sketches and images to serve directly as network specifications.

In this paper, we introduce \textit{GeNet}, a network engineering co-pilot,
and explore how to harness state-of-the-art tools to streamline network engineering workflows in enterprise environments.
\textit{GeNet} leverages multimodal \acp{LLM} to interpret network topology images and device configurations, enabling updates that align with user intents. 
\textit{GeNet} includes three key modules: the \emph{Topology Understanding} module, 
 the \emph{Intent Implementation} module 
and an automatic \ac{LLM}-based evaluation framework, capable of evaluating 
intent implementation quality with significant correlation to human evaluation. 

An example of a typical situation where \textit{GeNet} can assist even novice engineers:
\begin{example}[Adding Local PCs]
\label{ex:scenario}
The management has requested the addition of three workstations to a lab, but all ports in the respective switch are occupied. The engineer immediately inputs the above intent, along with an image of the lab's topology and the configurations of its raw components, without the need for special pre-processing. Based on the input, \textit{GeNet} understands that the Ethernet switch in the requested lab has run out of free ports. It suggests adding a switch in a daisy chain topology, connecting the additional workstations to it and outputs the updated configurations and topology.
\end{example}

We conduct a comprehensive assessment of \textit{GeNet}'s ability to handle a variety of intents, by evaluating its performance in a series of scenarios, characterized by varying complexities and input specification quality, formulated based on Cisco certification exercises and interviews with IT experts.
These intents include designing and updating network topologies and configurations.   

The main highlights of this research are as follows:
(1) We present \textit{GeNet}, a multimodal co-pilot designed for enterprise network engineers that facilitates network topology and configuration updates.
(2) We developed an automatic \ac{LLM}-based evaluation framework for \ac{IBN} intents, which shows a strong and significant correlation with human evaluations.
(3) We demonstrate that the capability of \acp{LLM} to interpret network topology images is significantly influenced by the quality of the inputs or any missing information. 
However, regardless of such input issues, \textit{GeNet}'s ability to successfully implement intents isn't significantly affected.


\section{Background and Related Work}\label{sec:background_relate-work}
\label{sec:related}

\ac{IBN} is a common automation approach that introduces an abstraction layer connecting high-level network requirements 
with low-level device configurations. 
IBN emphasizes the desired 
outcomes of network services rather than describing their implementation~\cite{leivadeas2022survey}.
While in traditional \ac{IBN} systems users have to develop proficiency in using predefined syntax~\cite{beckett2016don,ooi2023intent}, \ac{NLP}-based approaches offer greater freedom and flexibility~\cite{wang2024netconfeval}.


Recent \ac{IBN} studies examined the use of \acp{LLM} to streamline such processes.
Furthermore, based on the success of programming assistants such as GitHub Copilot, 
researchers are developing co-pilots capable of facilitating network engineering tasks. For instance, Zhao et al.~\cite{zhao2023confpilot} presented a framework that generates configurations given natural language intents, utilizing an \ac{LLM} and \ac{RAG}~\cite{lewis2020retrieval}. 

Several recent studies have introduced evaluation frameworks to assess \acp{LLM}' network reasoning abilities,
comparing popular \acp{LLM} on networking  tasks~\cite{wang2024netconfeval,donadel2024llms,miao2023empirical}. 
Most 
point out that GPT-4 
outperformed other popular models; 
hence, our research relies on GPT-4.
While these studies offer extensive evaluations of \acp{LLM} in networking, they have several limitations. 
Primarily, they focus on simplified and impractical tasks, such as generating small configuration snippets from scratch or translating informal specifications into formal ones.
In contrast, we demonstrate the ability to tackle realistic and diverse \ac{IBN} tasks, which involve modifying existing configurations or topologies comprehensively. 
Also, we examine realistic networks 
rather than simple topologies with only a few devices.
Moreover, many have utilized JSON-like data structures to convey both 
initial specifications and final configurations. 
This approach can simplify processes and improve performance, but it may also be impractical because it requires additional processing of the inputs or the outputs. 

 While fine-tuning an \ac{LLM} could enhance its performance in networking tasks~\cite{chen2022software}, it requires specialized datasets that are rare in 
 this domain, as well as frequent retraining to adapt to new technologies and standards. 
 Therefore, in this paper, we explore GPT-4's ability to implement network intents without fine-tuning.

Furthermore, although Donadel et al.~\cite{donadel2024llms} reported that \acp{LLM} struggle generating topology images, Wang et al.~\cite{wang2024netconfeval} raised the possibility that \acp{LLM}' understanding of networks could be enhanced by using topology images or graphs as context.
In this research, we aim to explore and build upon these insights and to the best of our knowledge, this is the first attempt to leverage the visual modality by translating network topology images into 
textual descriptions. 
Additionally, we are the first to utilize topology paper sketches and evaluate how GPT-4 understands different variants of network topologies. 

Finally, contrary to our research, which focuses on comprehensive network topology design and configuration, most related studies have primarily concentrated on generating network configurations.
\textit{GeNet} harnesses the capabilities of \ac{IBN} and multimodal \acp{LLM} to deliver user-friendly, holistic assistance to network engineers based on natural language intents and topology images. 
For a 
survey on the use of \acp{LLM} for 
networking refer to~\cite{hang2024large}.




\section{The \textit{GeNet} Framework} \label{sec:genet-framework}

In this section, we describe \textit{GeNet}, a co-pilot designed to assist network engineers in updating network topologies and device configurations. \textit{GeNet} simplifies these tasks by harnessing the capabilities of GPT-4, a state-of-the-art \ac{LLM}, to provide relevant and effective configuration and topology recommendations, aligned with the user's intents, which we call \emph{\textit{GeNet} solutions}. 
\textit{GeNet} receives a natural language intent and network specifications, including the topology image and device configurations, and outputs  
corresponding 
solutions.
   \begin{figure}[b]
    \centering
    \includegraphics[width=0.45\textwidth]{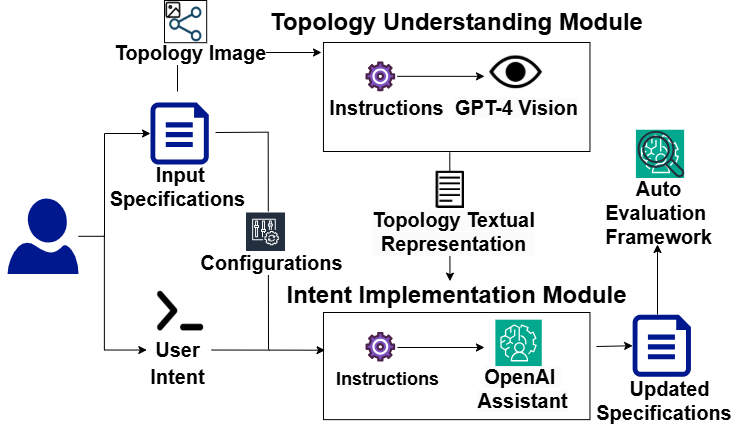}\caption{An overview of the \textit{GeNet} framework}
    \label{fig:architecture}
    \end{figure}
    
We designed \textit{GeNet} in a modular fashion, based on the 
assumption that \acp{LLM} demonstrate improved performance on smaller tasks~\cite{wang2024netconfeval}. 
\Cref{fig:architecture} presents the \textit{GeNet} framework and its key modules: the \emph{Topology Understanding}, 
the \emph{Intent Implementation}, 
and its automatic evaluation framework.
A description of the modules
and our research questions is provided below. 

\subsection{Topology Understanding Module}
\label{sec:topology-understanding-method}

In \textit{GeNet}'s first phase, the \emph{topology understanding} module generates a textual representation of the network topology based on the input image. 
This process uses GPT-4's \ac{VQA} capabilities\footnote{https://platform.openai.com/docs/guides/vision}.
Additionally, the \ac{LLM} receives a prompt containing precise guidelines, referred to as 'instructions' in \cref{fig:architecture}.

The model is assigned the role of a network topology analyst and instructed to describe the devices and edges in the image, including details such as icons and interface labels. 
To avoid redundancy, the instructions also specify to exclude unnecessary information, a tendency observed otherwise. 
These instructions, which serve as the \ac{LLM}'s system prompt, were refined through preliminary experiments to reduce output variability.

\subsection{Intent Implementation Module}
\label{sec:solver}
In \textit{GeNet}'s second phase, the \emph{intent implementation} module updates the network specifications so that they are aligned with the user's intent. 
This module utilizes OpenAI's Assistants API\footnote{https://platform.openai.com/docs/assistants/overview} 
which enables the creation of customized \acs{AI} assistants to perform complex tasks.
As before, the assistant is initialized with instructions, referred to as 'instructions' in~\cref{fig:architecture}.

The assistant is assigned the role of an expert in network topology and configuration, with precise guidelines regarding its responsibilities.
These guidelines include updating the textual representation of the topology to accurately reflect the user's intent, configuring newly added components,
and modifying existing device configurations as necessary. 
Furthermore, the assistant is instructed to explain all changes made during the process to help novice users to build expertise in the domain.

The assistant accesses the textual representation of the topology and the device configurations, performs the necessary updates, and provides the modified files along with corresponding explanations. 
To maintain the flexibility of \textit{GeNet}, we avoid enforcing a specific configuration language in the instructions. 
Additionally, GPT-4 is used without fine-tuning to evaluate its inherent capability to effectively implement topology and configuration updates.
Notably, \textit{GeNet} is not tailored for GPT-4 and 
allows for the \ac{LLM} engine to be easily replaced.



\subsection{Automatic \ac{LLM}-Based Evaluation}
\label{sec:llm_eval}
We developed an automatic \ac{LLM}-based evaluation framework for \ac{IBN} intent implementations. 
It is grounded on the hypothesis that it is easier for an \ac{LLM} to evaluate a \emph{\textit{GeNet} Solution} using predefined scoring keys than to generate a high-quality solution. 
Consequently, we expected the \ac{LLM} to demonstrate strong evaluation capabilities.
We implemented this framework using the Assistants API, providing the assistant with instructions about its evaluation task, intent-specific scoring keys (Section \ref{sec:fst-scrn}), and both the input and output network specifications.

For an \ac{LLM} to function as an effective 
evaluator, it has to
achieve a significant positive correlation with human evaluation.
To ensure this, 
we refined both the instructions and the intent-specific scoring keys in an iterative process (Sec. \ref{sec:exp-setup}). 
Each time, the \ac{LLM} was instructed to provide brief explanations for its grades, allowing us to verify its understanding 
and reasoning process and refine the prompts
whenever 
misunderstandings arose.
All instructions and scoring keys mentioned above are available on our Git repository\footnote{https://github.com/networkcopilot}.

\subsection{Research Questions}\label{sec:rq}
The experiments performed in this work aim to answer the following research questions: 

\begin{resq}\label{rq:images}
How effectively does GPT-4 interpret network topology images of varying quality? (\ref{sec: rq1_answer})
\end{resq} 

\begin{resq}\label{rq:phase4}
How effectively can a multimodal \ac{LLM}-based networking co-pilot handle network intents? (\ref{sec: rq2_answer})
\end{resq}

\section{Experiments}\label{sec:eval}

\subsection{Dataset}\label{sec:data}
GNS3 Vault\footnote{https://gns3vault.com} offers training labs featuring scenarios of varying difficulty levels designed to help network engineers prepare for certification exams. 
Building on this concept, we interviewed IT experts to create a diverse dataset of ten scenarios, each consisting of an intent and network specifications, including device configurations and topology images.
Five of them focus on topology updates (adding DMZ, adding DRA, Internet connectivity, adding local PCs, and adding communication servers), 
while the other five concentrate solely on configuration changes (IP traffic export, basic zone-based firewall, role-based CLI access, time-based access list, and IOS transparent firewall). 

\begin{figure}[h]
    \centering
    \includegraphics[width=0.47\textwidth]{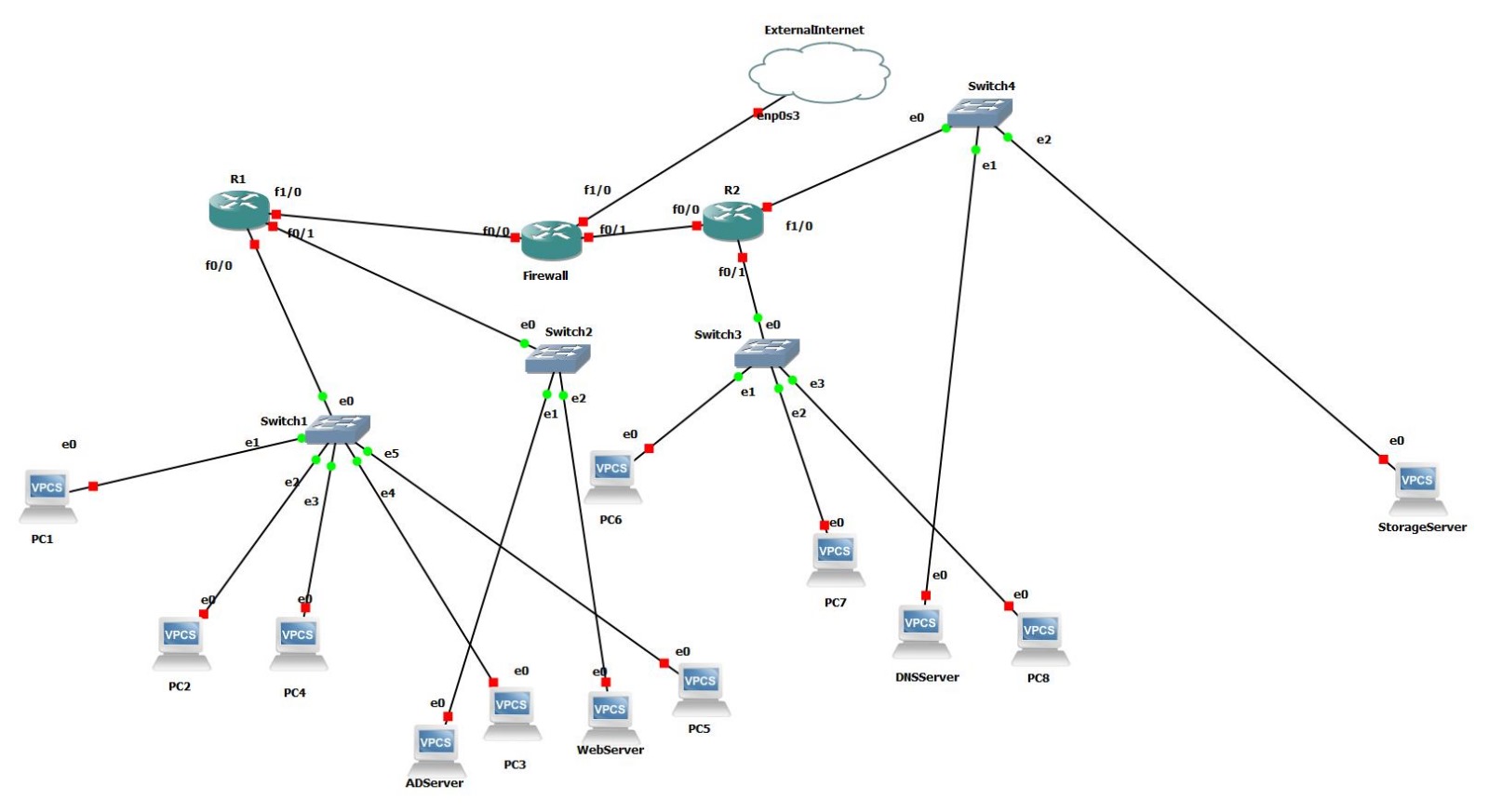}\caption{``Messy Layout" - GNS3 Example}
    \label{fig:exmp_messy_gns}
\end{figure}

To create the network specifications, which provide context for the user intent in each scenario, we simulated 
the scenarios in a network emulator, extracted the device configurations, and drew the topologies. 
For a thorough evaluation of GPT-4's ability to interpret network topology images, 
we created nine topology variants for each scenario. 
Each topology was created using three visualization formats: 
1) GNS3, a network emulator, 2) PowerPoint, a visualization software (without icons for devices), and 3) hand-drawn sketches.
Each format was used to create
three types of images: 1) \emph{Normal}, 2) \emph{No Labels on Edges}, and 3) \emph{Messy Layout}. 
The \emph{Normal} variant offers a balanced layout of the topology.
The \emph{No Labels on Edges} variant 
omits edge labels marking device interfaces.
The \emph{Messy Layout} variant, which highlights the characteristics of suboptimal visualizations, is created by randomly placing network components on the dashboard or paper and connecting them with edges (see \cref{fig:exmp_messy_gns}). 
We kept these variants human-readable, avoiding an incomprehensible representation.
The complete set of scenarios is publicly available in our Git repository.



\subsection{Experimental Setup}\label{sec:exp-setup}



Every topology image variant in each scenario was processed five times by \emph{GeNet}, resulting in a total of 450 \emph{\textit{GeNet} solutions} across 90 test cases.
Based on preliminary experiments, we employed GPT-4 Turbo with the temperature set to 1 (default value).


While human evaluation is generally considered more reliable, it is also labor-intensive and requires significant effort, which limits the scale of the experiments. 
To overcome this challenge, we manually evaluated \emph{GeNet solutions} from preliminary experiments using specific scoring criteria (Sec. \ref{sec:fst-scrn}). 
We used this evaluation to develop and validate our LLM-based evaluation framework (Sec. \ref{sec:llm_eval}), which allowed us to scale up our experiments effectively.
For this framework, we used the GPT4o model,
which was chosen based on preliminary experiments and its cost-effectiveness.
A temperature value of 0 was chosen to ensure the evaluation process is deterministic and aligned with the scoring keys.

\subsection{Solution Quality Criteria} \label{sec:quality}
\subsubsection{\Acf{TIUS}}\label{sec:topology-understanding-score}
The semi-structured output from the \emph{topology understanding} module was automatically compared to a manually constructed ground-truth representation for each scenario. 
The \ac{TIUS} metric quantifies the accuracy with which all visual elements in the topology image were described. 
To evaluate the \ac{LLM}'s accuracy in identifying elements of the topology, we measured the number of correctly identified elements of a specific type against the total number of elements of that type. 
The topology elements assessed include: nodes (e.g., routers), referred to as $N$; specific node labels, noted as $NL$; node icons, identified as $NI$; links between nodes, designated as $L$; and the connections of links to nodes (e.g., device interfaces), referred to as $LNL$.
The \Ac{TIUS} is calculated as a weighted sum of relative correct assessments:
\begin{equation}
    TIUS=0.3\cdot N+0.2\cdot NL+0.05\cdot NI+0.35\cdot L+0.1\cdot LNL
\end{equation}
The weights were assigned based on the relative importance of errors in different elements. 
This score is primarily used to assess the performance of the \emph{Topology Understanding} module.

\subsubsection{\Acf{IIS}} \label{sec:fst-scrn}
Since each scenario includes an intent and network specifications, we developed scenario-specific scoring keys to evaluate how effectively \textit{GeNet} implemented each intent. 
These scoring keys were initially formulated using the scenario solutions provided by GNS3 Vault for the configuration scenarios and insights from IT expert interviews for the topology scenarios. 
During the evaluation, additional scoring keys were introduced to account for \textit{GeNet}'s common errors and correct \emph{\textit{GeNet} solutions} that differed significantly from those provided by GNS3 Vault. 
When a new valid solution approach or a recurring mistake was identified, the scoring keys were updated, and all prior results were re-evaluated to ensure consistency.
Each non-compliance with a key results in a full or partial deduction of the specified points according to the severity of the non-compliance.
This scoring system is primarily used to assess \textit{GeNet}'s overall performance.
\begin{example}[Adding Local PCs]
\label{ex:scoring}
For the scenario presented as an example (\ref{ex:scenario}), the following scoring keys were suggested: 
1) Adding a switch to the topology: 20 points. 
2) Adding 3 PCs: 10 points per PC.
3) Correct subnet assignment: 3 points.
4) Configuring the new switch: 15 points.

\end{example}



\subsection{Results}
\label{sec:results}


During our manual evaluation of \emph{\textit{GeNet}'s Solutions} in the preliminary experiments, we found that out of 250 test cases, \textit{GeNet} successfully configured 79\% of the expected network devices. Additionally, \textit{GeNet} reconfigured 92\% of the devices from the input specifications that required modifications to comply with intents. 
Furthermore, it configured 72\% of the newly added components that were not included in the initial input. 
This may suggest that configuring network devices from scratch poses a greater challenge for \textit{GeNet}.
In addition, on average, the \emph{Topology Understanding} module processed topology images in 24 seconds each, while the \emph{Intent Implementation} module executed intents in 57 seconds. 
This results in a total average processing time of 81 seconds per scenario. 


\subsubsection{Automatic \ac{LLM}-based Evaluation} \label{sec:llm_eval_res}
\begin{figure}[t]
    \centering
    \includegraphics[width=0.40\linewidth]{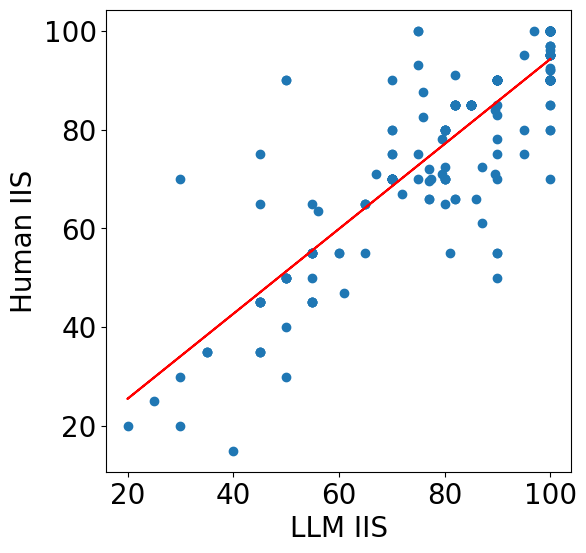}\caption{The correlation between human and LLM \ac{IIS}}
    \label{fig: human vs LLM}
\end{figure}

Various approaches and prompts were tested to optimize the correlation between the \ac{LLM} and human evaluation.
Initially, we instructed the \ac{LLM} to provide only a final evaluation score of a \emph{GeNet solution} based on the scoring keys, without further explanations, however, this approach resulted in a low correlation.
We observed that the \ac{LLM} often assigned general grades (e.g., 85) to many solutions of varying quality without any rationale. 
We believe this issue stemmed from the complexity of the evaluation task, which requires comprehension of the grading instructions, scoring keys, and intent of the scenario in the context of both the original and updated specifications. 
To improve performance, the evaluation was split into smaller sub-tasks by having the \ac{LLM} evaluate each scoring key separately.
This increased the correlation, though insufficiently.
Next, we applied the Chain-of-Though approach \cite{wei2022chain} by prompting the \ac{LLM} to provide brief explanations for each scoring key, potentially enhancing its understanding of the task at hand.
This led to a significant improvement in alignment with human evaluations, with the Spearman correlation rising to $0.847$ $(p=4.11 \times 10^{-67})$, indicating strong agreement (see \cref{fig: human vs LLM}).
We believe that asking the \ac{LLM} to explain its grading for each key helped it focus on the relevant parts of the \emph{GeNet Solution}, rather than trying to reason about the solution as a whole.


\subsubsection{Full experiment}
\begin{figure}[h]
    \centering
    \includegraphics[width=0.8\linewidth]{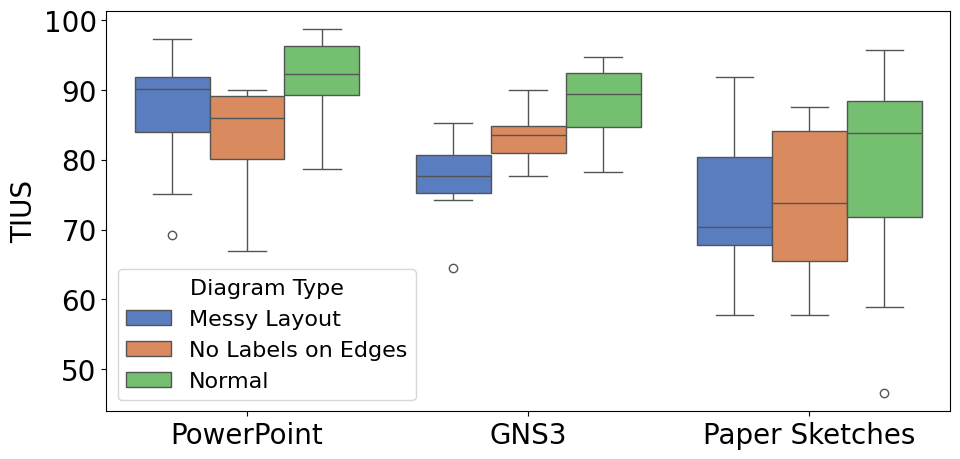}\caption{ \ac{TIUS} by topology format and type}   
    \label{fig: TIUS_boxplots}
\end{figure}

\begin{figure}[t]
 \centering
\begin{subfigure}{.40\linewidth}
    \centering
    \includegraphics[width=1\linewidth]{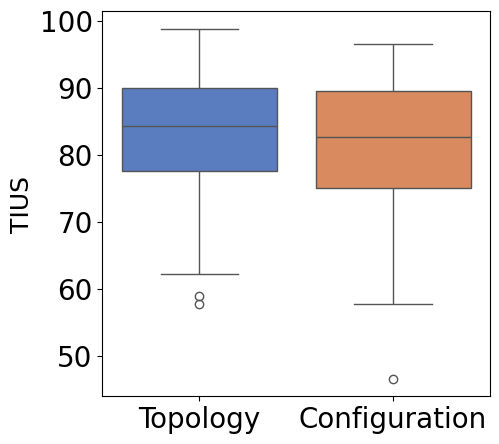}  
    \caption{}
    \label{fig:TIUS vs type - full}
\end{subfigure}
\begin{subfigure}{.40\linewidth}
    \centering
    \includegraphics[width=1\linewidth]{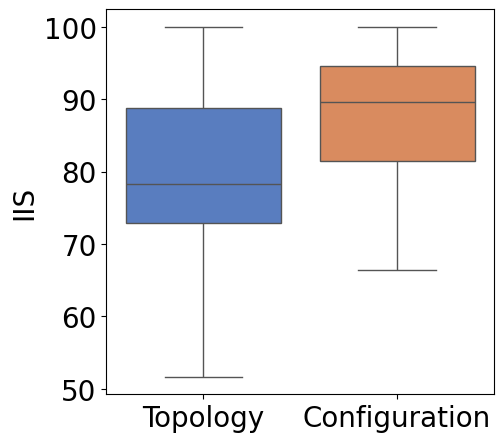}  
    \caption{}
    \label{fig:IIS vs type}
\end{subfigure}
\caption{Scenario type effect on the \ac{TIUS} (a) and \ac{IIS} (b)}
\end{figure}

\paragraph{Topology image understanding (\ref{rq:images})} 
\label{sec: rq1_answer}
Overall, the \emph{topology understanding} module achieved high scores in topology understanding, with a mean \ac{TIUS} of 0.73 for the \emph{Messy Layout-Paper Sketch} variant, which is considered as the most difficult to comprehend (see \cref{fig: TIUS_boxplots}). 
This result confirms the potential of multimodal \ac{LLM}-based co-pilots to effectively utilize network topology images, as raised by \cite{wang2024netconfeval}. 
Additionally, as illustrated in \cref{fig:TIUS vs type - full} and based on a mean equality t-test, the \ac{TIUS} does not significantly depend on the scenario type.

To evaluate the impact of the scenarios, visualization formats, and image types on the \ac{TIUS}, we measured their \ac{IG}. 
The scenarios alone accounted for most of the variance in the \ac{TIUS}, with an \ac{IG} value of 2.25. 
This can be attributed to the differing topology complexity in each scenario. 
For instance, the topology in the "Adding DRA" scenario is considerably more complex than that in the "Time-Based Access List" scenario. 
Moreover, as expected, the visualization formats significantly influenced the \ac{TIUS}. 
As illustrated in \cref{fig: TIUS_boxplots}, and based on a mean equality t-tests, PowerPoint produced significantly higher \ac{TIUS} scores compared to images created with GNS3 or sketched on paper. 
It is not surprising that the \emph{Paper Sketches} yielded the lowest \ac{TIUS}, as this format lacks the clarity and structure of digital visualizations. 
However, it is quite surprising that PowerPoint, a general visualization tool, resulted in higher \ac{TIUS} than GNS3, a specialized networking tool. 
This might suggest that topologies created with PowerPoint and not containing device icons are more comprehensible for GPT-4.
Furthermore, the image types had a notable impact on the \ac{TIUS}. 
As anticipated, the \emph{Normal} type resulted in a significantly higher \ac{TIUS} compared to the \emph{Messy Layout} and \emph{No Labels on Edges} types. 
However, there was no significant difference in \ac{TIUS} between the latter two. 
One possible explanation is that the \emph{Messy Layout} type inherently reduces \ac{TIUS} due to its suboptimal visualization characteristics, while the \emph{No Labels on Edges} type lacks critical information that could aid in accurately interpreting the topologies. 
Interestingly, in some instances, the \emph{topology understanding} module was able to perfectly comprehend images of the \emph{Messy Layout} type, achieving a \ac{TIUS} of 100. 
This suggests that \textit{GeNet} can interpret low-quality specifications correctly.

\paragraph{Intent implementation (\ref{rq:phase4})}
\label{sec: rq2_answer}
On average, \textit{GeNet} performed better in the configuration scenarios than in the topology scenarios (\cref{fig:IIS vs type}). 
A t-test and Levene's test for mean and variance equality, respectively, indicated significant differences in the \ac{IIS} between the two scenario types. 
This suggests that the topology update tasks in our dataset are consistently more challenging for \textit{GeNet} than the configuration ones. 
One possible explanation 
is that our configuration scenarios typically involve updating a single device, whereas our topology scenarios require the configuration of multiple devices, hence 
demand a longer and more complex thought process.

Another potential explanation is that our configuration scenarios do not require devices to be configured from scratch. 
In contrast, our topology scenarios typically demand adding new devices and configuring them.
It is well established that \acp{LLM} tend to generate more accurate answers based on a given context or template, as opposed to generating answers from scratch. 
Thus, configuration scenarios, which mainly involve updating existing device configurations, may be easier for the \ac{LLM} to handle.
In fact, the observations of our preliminary experiment support this assumption (Sec. \ref{sec:results}). 

\begin{figure}[h]
    \centering
    \includegraphics[width=0.8\linewidth]{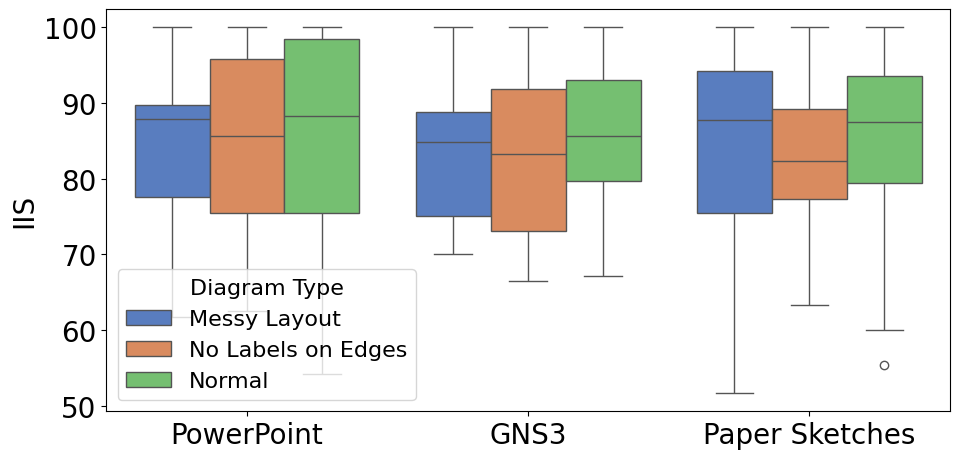}\caption{\ac{IIS} by topology format and type}   
    \label{fig: IIS_boxplots}
\end{figure}

To deeply investigate the influence of the visualization formats and types on the \ac{IIS}, we conducted a series of pairwise statistical tests to assess mean and variance equality across different format-type pairs (e.g., GNS3-Messy Layout vs. PowerPoint-Normal). 
These tests showed insignificant differences in both mean and variance across all pairs regarding the IIS. 
As illustrated in \cref{fig: IIS_boxplots}, there is no consistent relationship between these pairs in terms of the \ac{IIS}. 
Additionally, the various visualization formats and types exhibited the lowest \acf{IG} concerning the IIS, compared to the \ac{TIUS}, the scenario, and the scenario type variables. 
This underscores \textit{GeNet}'s effectiveness in handling network specifications of varying quality (e.g., messy layouts) or with missing information (e.g., interface labels).

\section{Conclusion and Future Work}\label{sec:conclusion}
In this paper, we introduce \textit{GeNet}, a multimodal co-pilot designed for enterprise network engineers. \textit{GeNet} is an innovative framework that utilizes an \ac{LLM} to streamline network design workflows by updating network topologies and device configurations based on user intents. 
Our results demonstrate that \textit{GeNet} can effectively interpret topology images, highlighting the potential of multimodal \ac{LLM}-based co-pilots in leveraging such visual data. 
This can significantly reduce engineers' efforts and accelerate network design processes.
Furthermore, \textit{GeNet} shows the ability to successfully handle various intents, 
even with incomplete or low-quality input specifications, 
in an average processing time of 81 seconds.


\begin{acronym}
\acro{IBN}{intent-based networking}
\acro{NLP}{natural language processing}
\acro{LLM}{large language model}
\acro{SMT}{satisfiability modulo theories}
\acro{AI}{artificial intelligence}
\acro{RNN}{recurrent neural network} 
\acro{DL}{deep learning}
\acro{CV}{computer vision}
\acro{VQA}{visual question answering}
\acro{SDN}{software-defined network}
\acro{GNN}{graph neural network}
\acro{RAG}{retrieval augmented generation}
\acro{IG}{information gain}
\acro{TIUS}{topology image understanding score}
\acro{FSS}{fast screening score}
\acro{EVSS}{emulator validated screening score}
\acro{S3}{solution screening score}
\acro{S3-Z}{standardized solution screening score}
\acro{IIS}{Intent Implementation Score}
\end{acronym}

\bibliographystyle{IEEEtran}
\bibliography{main}

\end{document}